\documentclass[10pt,pra,aps,twocolumn,superscriptaddress,floatfix,notitlepage,nofootinbib]{revtex4-2}
\usepackage[latin1]{inputenc}
\usepackage{hyperref}
\hypersetup{
    colorlinks=true,
    linkcolor=blue,
    filecolor=magenta,      
    urlcolor=cyan,
    citecolor=blue,
    pdftitle={Correlated Quantum Sensing at the Seemingly Classical Limit},
    pdfpagemode=FullScreen,
    }

\usepackage{mathrsfs,dsfont}
\usepackage{amsmath}
\usepackage{varwidth}
\usepackage{empheq}
\usepackage{braket}
\usepackage{amsfonts,natbib}
\usepackage{amssymb}
\usepackage{bm}
\usepackage[pdftex]{graphicx}
\usepackage[normalem]{ulem}

\date{\today}

\usepackage{graphicx}
\begin{document}
\title{Correlated Quantum Sensing at the Seemingly Classical Limit}
\author{K. P. Athulya}
\email{athulyakp@tifrh.res.in}
\affiliation{Tata Institute of Fundamental Research Hyderabad, 36/P, Gopanpally Village, Serilingampally Mandal, Hyderabad, Telangana 500046, India}
\author{Sreenath K. Manikandan}
\email{skm@tifrh.res.in}
\affiliation{Tata Institute of Fundamental Research Hyderabad, 36/P, Gopanpally Village, Serilingampally Mandal, Hyderabad, Telangana 500046, India}
\date{\today}
 \begin{abstract} It is a difficult task to detect the indivisible quanta of weakly interacting radiation fields, and even more challenging to
probe their quantum statistics. Nevertheless, if barely functional high-quality resonant detectors are feasible for weakly interacting radiation fields, they do come with certain statistical advantages to probe quantum effects at the seemingly classical limit of a large number of quanta of the incoming radiation field. 
 We present correlated counting, homodyne, and heterodyne detection strategies using high-quality resonant quantum harmonic detectors operating at this limit, initialized in bolometry-inspired zero-mean preparations such as thermal states. We compare the bolometric regime of
 good resonant harmonic detectors in quantum optics to the bolometric regime of barely functional resonant mass quadrupole oscillators as detectors for quantum gravity. Simple statistical tests are proposed using symmetric correlators for two and three such barely functional resonant mass detectors that could reveal the complementary quantum noise characteristics of gravitons in tabletop experiments. 
 \end{abstract}
 	\maketitle
\noindent\textit{Introduction.---}
Quantum mechanics originated from invoking purely statistical and thermodynamic considerations for light and light-matter interactions,
 in rather finesse experiments, such as measurements of the black body radiation spectrum, performed at finite temperature~\cite{Planck:1901oar}. Explaining the low-temperature specific heat of solids was another major early victory for quantum theory~\cite{Einstein}. Although these early triumphs involved experiments at finite temperature~\cite{Pais:1979vn}, and a large number of radiation quanta, the modern frontiers of light-matter interactions, in particular, quantum sensing for fundamental physics, rely heavily on sub-milli-Kelvin temperature physics, capabilities for quantum ground state cooling, and resources such as quantum entanglement and squeezing~\cite{Bose:2024nhv,Giazotto:2006zz,LIGOScientific:2013pcc,LIGOScientific:2011imx,Bass:2023hoi},  often involving only a few quanta of the field~\cite{Pirandola:2018osg,Hadfield:2009gyi,Hadfield:23,Oxborrow01052005}. 

In these regards, an inspiring recent progress on the experimental frontier is the renewed interest in bolometric detectors~\cite{langley1880bolometer,langley1881bolometer,Cabrera:1984rr,Drukier:1984vhf,Krauss:1985aaa}, in quantum optics~\cite{Sarkar:2025byz,Catto:2023ouc,Kokkoniemi:2020mou,Karimi:2024cog,Charaev:2024ptd}. Here, thermodynamic signatures, primarily temperature changes in a low specific heat sample, have been used as a probe to detect radiation fields, even suggesting prospects for probing light at the single quantum level~\cite{Catto:2023ouc}. However, the detection efficiencies are not expected to be substantial enough to be used as a direct quantum sensing strategy with high statistical significance. This trade-off is also evident from the comparative studies of the energetic cost of using thermal, coherent, and quantum light for quantum measurements~\cite{Linpeng:2022qxi} and control operations~\cite{Simbierowicz:2024qki}.  
While typical photodetectors have close to unit quantum to click conversion efficiency enabling them to respond at very weak intensities of the incoming radiation field, recent works involving gravitational radiation~\cite{Tobar:2023ksi,Manikandan:2025ykr,Manikandan:2025hlz,Manikandan:2025lfx,Manikandan:2025qgv,Carney:2023nzz,Domcke:2024mfu,Carney:2024dsj,Parikh:2020kfh,Parikh:2020nrd} motivate inefficient probes for radiation fields of a fundamentally different kind, which only respond at the seemingly classical limit of a large number of incoming quanta, $\langle N\rangle=\langle a^\dagger a\rangle\gg 1$. These detectors are assumed to couple very weakly to the radiation fields of interest, but their physical parameters are optimized such that if they were to be initialized in their quantum ground states, single quanta of the incoming field can, in principle, be detected with high fidelity using quantum sensing. Given that $\gamma_0$ is the spontaneous emission rate for the detector to spontaneously emit a quantum of the radiation field to be detected, $\tau$ to be the detection window, and $\langle N\rangle$ be the average number of quanta in the incoming field, a very weakly coupled detector operates in the limits $\eta\equiv\gamma_0\tau\ll 1$ and $\langle N\rangle\gg 1$, nevertheless the stimulated absorption probability, $\eta\langle N\rangle$ is finite and of the order of unity.  In comparison, a conventional single photon detector operates in the regime $\eta\sim 1$ such that the functional detector regime of $\eta\langle N\rangle\sim 1$ can already be achieved for $\langle N\rangle\sim 1$.   

    \begin{figure*}[tb]
\includegraphics[width=\linewidth]{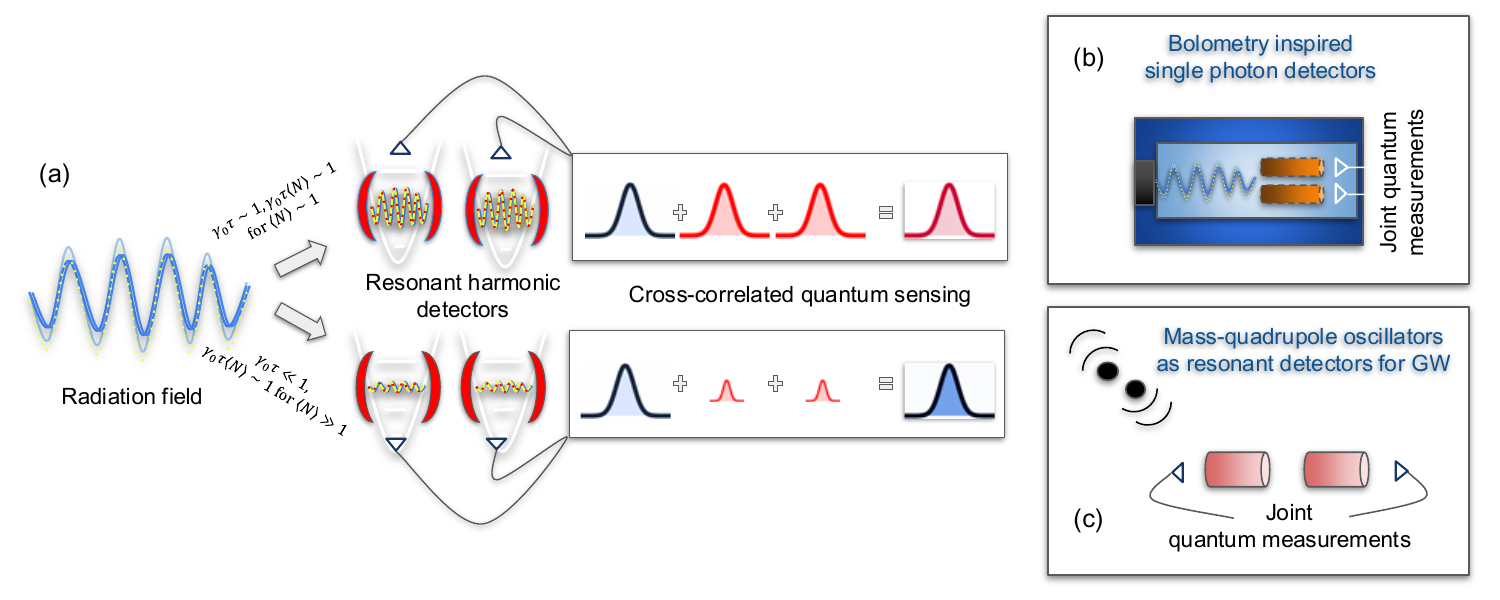}
   \caption{(a) Cross-correlated quantum sensing using two or more high-quality, but barely functional resonant quantum harmonic detectors in the bolometric regime can offer a statistical advantage for probing the coherence properties of weak fields in the seemingly classical limit of large $\langle N\rangle$. We compare (b) the bolometric regime of good resonant harmonic detectors for probing quantum aspects of light to (c) the bolometric regime of barely functional resonant mass detectors for probing the quantum nature of gravitational radiation. The comparison is made in the functional detector regime, i.e., $\gamma_0\tau\langle N\rangle \sim 1$ for both cases.}
    \label{fig:Model}
\end{figure*} 

The two fundamentally different inefficient detector regimes described above jointly motivate us to consider very weakly coupled resonant detectors initialized in zero-mean states, such as thermal states, as ``barely functional" resonant detectors in the bolometric regime.   We first extend the radiation counting statistics arguments to high-quality resonant detectors in thermal states, which also shows that they will be very inefficient to be used as direct quantum sensors (although other zero mean states, such as Fock states, can offer advantages). Hence, we focus on correlated quantum sensing strategies using such barely functional detectors, with the motivation to probe complementary quantum noise characteristics of radiation fields in the seemingly classical limit of large $\langle N\rangle$.  
This limit of barely functional detectors is well motivated based on various resonant detection schemes for the indivisible quanta of weak fields, including fields that are yet to be detected~\cite{Adams:2022pbo}, or fields those have been detected classically~\cite{LIGOScientific:2016aoc}, but their quantum properties are yet to be verified~\cite{Manikandan:2025ykr,Tobar:2023ksi,Manikandan:2025lfx,Manikandan:2025qgv,Dyson:2013hbl,Carney:2023nzz,Parikh:2020kfh,Parikh:2020nrd,Shenderov:2024rup,Toccacelo:2026hcz,Carney:2024dsj,Loughlin:2025rih,Tobar:2024bjr,Dyson:2013hbl}. These appear to satisfy the above conditions naturally because a major challenge in detection is that their coupling to matter or light is fundamentally weak. High-quality detectors may be within reach in many of these frontiers, although ground-state cooling can be a major challenge. A motivating example for a weakly coupled, yet potentially functional detector that checks all the above is a resonant mass-quadrupole oscillator as a quantum harmonic detector for individual  gravitons~\cite{Tobar:2023ksi}. They are acoustic bar resonators of the type Weber proposed~\cite{Weber:1960zz} to detect classical gravitational radiation, but it was recently suggested that such acoustic bar resonators functioning as quantized mass quadrupole oscillators with the measurements of quantum jumps, could function as counting detectors of the quanta of the gravitational radiation field~\cite{Tobar:2023ksi}. The corresponding conversion efficiency of graviton detectors is fundamentally tiny, 
$\eta\sim 10^{-36}\ll 1$~\cite{Manikandan:2025ykr,Tobar:2023ksi,Shenderov:2024rup,Manikandan:2025hlz}. These imply that it would take an incoming radiation field of as many as $10^{36}$ gravitons for the graviton detector to register a click, such that the click probability, $P_{0\rightarrow 1}\sim O(1)$~\cite{Tobar:2023ksi,Shenderov:2024rup}. The possibility of detecting gravitons through graviton-to-photon conversion also shares similar characteristics~\cite{Carney:2023nzz,Carney:2024dsj}. The seemingly classical gravitational radiation fields measured at the Laser Interferometer Gravitational-Wave Observatory (LIGO) contain as many gravitons in their quantum description, and therefore could produce a measurable response in the resonant mass detectors too. 
Such ground-state cooled detectors have also been discussed as capable detectors for probing the noise of gravitons through tests of the coherent state hypothesis~\cite{Manikandan:2025ykr,Manikandan:2025qgv,Manikandan:2025hlz,Toccacelo:2026hcz}, including via cross-correlations between quantum ground state cooled detectors~\cite{Manikandan:2025lfx,Bilinskaya:2026ryc,Toccacelo:2026hcz}. However, largely the emphasis has been so far on detectors initialized in their quantum ground states. The weak coupling is thought to be a major challenge~\cite{Carney:2024dsj,Carney:2023nzz}, but the bolometric regime we consider benefits from it, and suggests that bolometry-inspired thermal initial preparations of resonant harmonic detectors could alleviate some of the quantum ground state cooling requirements to test complementary quantum noise characteristics for gravitons as well, through correlated number and phase quantum measurements. 

 Our findings have synergies with earlier observations that homodyne cross-correlation measurements using a weak local oscillator can have certain advantages to probe quantum statistics of the sub-Poissonian nature, as in fluorescence~\cite{Yuen:83,Abbas:83,Schumaker:84,WernerHomodyne,Vogel2}, also in Hong-Ou-Mandel type measurements involving losses~\cite{DuttaGupta:14}.  In comparison, we mainly focus on counting measurements with two and three-detector configurations that probe super-Poissonian characteristics, while also discussing the homodyne and heterodyne implementations in the bolometric regime as additional probes for sub-Poissonian quantum characteristics of the weak field of interest. Importantly, the advantage vanishes at strong coupling with the field of interest, making it a unique advantage in the weakly coupled bolometric detector regime. The comparison is made assuming $\gamma_0\tau\langle N\rangle \sim 1$ for both limits (see Fig.~\ref{fig:Model}.). Our interest is also in the seemingly classical limit of the field, and the extreme limit of barely functional detectors relevant for probing the quantum noise of gravitons, i.e., $\gamma_0\tau\rightarrow 0, \langle N\rangle \rightarrow \infty:\gamma_0\tau\langle N\rangle \rightarrow 1$.

\noindent\textit{The detector model.---} We generalize a physically motivated model of a quantum harmonic oscillator as the resonant detector. This model is particularly pleasing with applications extending to a wide class of radiation detection problems, as it yields Mandel's counting statistics in a straightforward manner for detectors initialized in the vacuum state.   
While incoming radiation fields typically involve multiple modes of the incoming field, one could make an effective single-mode approximation for the field, with a revised coupling rate that can be understood as the square root of the spontaneous emission rate, $\sqrt{\gamma_0} $, for the detector to emit a quantum of the radiation field we intend to detect. This is described in the interaction picture by the following Hamiltonian~\cite{Manikandan:2025ykr}, 
\begin{equation}
    H_{I}(t) = \hbar\sqrt{\gamma_{0}} \left[d(t) a^\dagger + d^\dagger(t)a \right]
\end{equation}
where $a$ denotes the radiation field mode and $d(t)$ represents the detector mode. Relevance to bolometry-inspired detection strategies in quantum optics is that the standard dipole interaction Hamiltonian $H_I(t) = -\textbf{d}.\textbf{E}(t)$ can also be brought to the above form, where $\gamma_0$ will be the spontaneous emission rate for photons derived from the Wigner-Weisskopf theory~\cite{Weisskopf:1930au}. Also see Appendix.~\ref{App_bolo}. The use of this interaction to describe ground-state cooled resonant mass detectors is discussed in Refs.~\cite{Manikandan:2025ykr,Manikandan:2025hlz,Manikandan:2025lfx,Manikandan:2025qgv}.  
Assuming a Markovian resetting of the detector modes, $ \left[ d(t'), d^\dagger(t'')\right] = \delta(t' - t'')$
and upon redefining an effective detector mode over the observation window, $    b = \frac{1}{\sqrt{\tau}} \int_t^{t+\tau} d(t') dt'
$
such that $[b, b^\dagger] = 1$ is obeyed, the time-evolution operator reduces to the following form~\cite{Manikandan:2025ykr},
\begin{equation}
    U_{I} = \exp \big[-i\sqrt{\gamma_{0} \tau} ( ba^\dagger + b^\dagger a) \big],
\end{equation}
which corresponds to an effective interaction Hamiltonian obeying $\tau H_{\text{eff}} = \hbar\sqrt{\gamma_{0} \tau} ( ba^\dagger + b^\dagger a).$ This indeed describes an energy and number-conserving resonant interaction between the field and the detector. For the graviton detector proposed in Ref.~\cite{Tobar:2023ksi}, $\gamma_0= 8GML^2\omega^4/(\pi^4c^5)\approx 10^{-33}$Hz at kilo-Hz frequencies. Assuming a detection window $\tau\sim 1$ ms, yields the estimate $\eta \sim 10^{-36}\ll 1,$ for the detector's graviton to click conversion efficiency, demonstrating that it satisfies our extreme weak coupling requirement.

\noindent\textit{The counting statistics for generic detector preparations.---}We begin by considering a single detector and radiation field in the initial state,
\begin{equation}
\rho(0) = \rho_{F}(0) \otimes \rho_{D}(0).
\end{equation}
Using the Glauber-Sudarshan $P$ representation for both the field and detector~\cite{Sudarshan:1963ts,Glauber:1963tx}, the time-evolved state of the field and detector after the interaction is, 
\begin{equation}
\rho' 
= \int d^2\alpha \, d^2\beta \;
P_{F}(\alpha)\, P_{D}(\beta)\;
 \ket{\alpha'}\bra{\alpha'} \otimes \ket{\beta'}\bra{\beta'}.
\end{equation}
where (equivalent to beam-splitter relations~\cite{Gerry_Knight_2004}),
\begin{align}
\alpha' &= \alpha\cos(\sqrt{\gamma_0 \tau}) - i\beta\sin(\sqrt{\gamma_0 \tau}),\\
\beta'  &= \beta\cos(\sqrt{\gamma_0 \tau}) - i\alpha\sin(\sqrt{\gamma_0 \tau}).
\end{align}

The probability $P_n$ that the detector registers $n$ quanta is given by projecting the detector onto the number (Fock) basis $\{\ket{n}\}$ and tracing over the field, given by,
\begin{eqnarray}
P_{n} &=& 
\frac{1}{n!}
\int d^2\alpha d^2\beta P_{F}(\alpha) P_{D}(\beta)\nonumber\\
&\times&|\beta\cos\theta- i\alpha\sin\theta|^{2n}
e^{-|\beta\cos\theta - i\alpha\sin\theta|^2}.\label{EqStat1}
\end{eqnarray}
where $\theta = \sqrt{\gamma_0 \tau}$. 
By assuming zero-mean states of the detector, applying the small-angle approximation in the weak coupling or short time limit, the average number of detected quanta and the corresponding number fluctuations reduce to (see Appendix.~\ref{Appx-A} for details), 
\begin{equation}
    \bar{n} \simeq \gamma_0 \tau \langle a^\dagger a \rangle + \langle b^\dagger b\rangle (1-\gamma_0\tau),
\end{equation}
and,
\begin{equation}
\begin{split}
\Delta n^2
&\simeq\langle b^\dagger b\rangle
 + \langle b^\dagger b\rangle Q_b\\
&\quad
 +  \gamma_0 \tau (2\langle a^\dagger a \rangle \langle b^\dagger b \rangle-2\langle b^\dagger b\rangle Q_b+ \langle a^\dagger a\rangle-\langle b^\dagger b\rangle)\\
&\quad+ (\gamma_0 \tau)^2\,[\langle a^\dagger a\rangle Q_a+5\langle b^\dagger b\rangle Q_b/3-8\langle a^\dagger a\rangle\langle b^\dagger b\rangle/3\\&\quad-(\langle a^\dagger a\rangle-\langle b^\dagger b\rangle)/3].
\end{split}
\end{equation}
where 
$Q_{a,b}=\left[(\Delta N_{a,b})^2-\langle N_{a,b}\rangle\right]/\langle N_{a,b}\rangle$ is the Mandel Q parameter of the radiation field and detector, respectively. In the barely functional detector regime, the value of the term $\gamma_0 \tau\ll 1$, however, $\gamma_0 \tau \langle a^\dagger a\rangle$ could still be of the order of unity. In this case, 
the average counts would approximately be the order of $\sim\langle N_b\rangle+1$, which is not distinguishable from $\langle N_{b}\rangle$ in the large $\langle N_b\rangle$ limit. Although for small $\langle N_b\rangle$, the difference could be substantial, which is the typical regime of resonant detection with near-ground state cooling. 

As an example of a zero-mean initial state for the detector that differs from vacuum, we may also consider a Fock (number) state, $|n^D\rangle$. In this case $Q_b = -1$, which suggest that the detector variance becomes,
\begin{eqnarray}
    &&(\Delta n)^2 = \gamma_0\tau\{n^D+\langle a^\dagger a\rangle \left[1+2n^D\right]\}\nonumber\\
    &&+\frac{(\gamma_0\tau)^2}{3} (3Q_a\langle a^\dagger a\rangle -8\langle a^\dagger a\rangle n^D-4n^D-\langle a^\dagger a\rangle).\nonumber\\&&
\end{eqnarray}
We see that for finite $n^D$, using observable fluctuations on a barely functional detector ($\gamma_0\tau\ll 1,\gamma_0\tau\langle a^\dagger a\rangle\sim 1$), a thermal state of the field ($Q\sim \langle a^\dagger a\rangle$) or a highly squeezed vacuum state of the field ($Q\sim 2\langle a^\dagger a\rangle$) can be discriminated from a coherent state for which $Q_a = 0$. This indeed indicates that there are certain advantages of highly quantum mechanical initial preparations of a detector, although such preparations can be hard to achieve in practice, especially for graviton detectors. 

We now consider easy-to-prepare zero-mean states such as thermal states for which $Q\sim \langle N\rangle$. For this case, considering weakly coupled radiation fields that are detectable ($\gamma_0 \tau \langle a^\dagger a\rangle\sim 1$), we can approximate, $(\Delta n)^2\sim \bar{n}+\langle N_b\rangle^2+2\langle N_b\rangle +\gamma_0\tau (Q_a-2\langle N_b\rangle^2)$. We see that the excess noise will be dominated by the detector's own noise, ruling out the possibility of statistical discrimination between different quantum states of the radiation field, using the excess noise observable in a single detector. It is also shown in Appendix.~\ref{App_bolo} that for incoming radiation field in a thermal state, the detector upon statistical average evolves into a new thermal state at a new effective temperature $ \hbar\omega/(k_{B}T_{\rm{eff}}) = \ln\left[1+(\bar{n}_D \cos^2\theta + \bar{n}_F \sin^2\theta)^{-1}\right]$, measurements of which correspond to bolometric detection. 

These observations indeed go with the usual agreement that thermal detectors are not quantum efficient. However, given that we assume detectors are of high quality, they will not thermalise to ambient temperature quickly, and contain extractable quantum information in terms of deviations from a thermal state, which, however maybe challenging to obtain. We now proceed to discuss correlated quantum sensing strategies that would make this additional quantum information extractable. Interestingly, we show that this would be possible only if the radiation field is coupled very weakly to a resonant detector, i.e., in the barely functional bolometric detector regime. 

\noindent\textit{Cross-correlated quantum sensing.---}
We now consider more than one resonant quantum harmonic detectors, which interact simultaneously with the incoming radiation field. The interaction Hamiltonian for the radiation field and the detectors is given by $H_I \tau = \tau\sum_i H_{I,i},$ where,
\begin{equation}
\tau H_{I,i} =\hbar\sqrt{\gamma_0\tau}
\left(a^\dagger b_i+b_i^\dagger a \right),~~i=1,2,..k., 
\end{equation}
Though we will restrict to $k=2,3$ in this work for concreteness and simplicity, our results are generalizable.
 
We consider arbitrary initial states of the field and the detectors, subsequently restricting our attention to the case of detectors prepared in thermal states. 
Following the interaction between the detectors and the radiation field, the relevant joint probability distribution of measurable counts in the Glauber-Sudarshan $P$ representation is given by,
\begin{align}
P(n_1,n_2)
&=\int d^2\alpha d^2\beta_1 d^2\beta_2
P_{F}(\alpha)P_{1}(\beta_1)P_{2}(\beta_2)\nonumber\\
&\quad\times
\frac{|\beta_1'|^{2n_1}}{n_1!}e^{-|\beta_1'|^2}
\frac{|\beta_2'|^{2n_2}}{n_2!}e^{-|\beta_2'|^2},
\end{align}
where $\beta_1',\beta_2'$ are estimated through  a normal mode analysis presented in Appendix.~\ref{Appx-B-two_detector}. From this, we can calculate the symmetric correlator between two detectors prepared in initial states such as thermal states or Fock states for which $\langle b_i\rangle, \langle b_i^2\rangle, \langle b^\dagger_i b^2_i\rangle$ etc vanish,
\begin{eqnarray}
   \kappa_2&\equiv& \langle (N_{1}-\langle N_1\rangle)(N_{2}-\langle N_2\rangle)\rangle\nonumber\\&=&  \langle N_1 N_2 \rangle - \langle N_1 \rangle \langle N_2 \rangle=\frac{1}{4} \sin^4(\sqrt{2\gamma_0 \tau})\nonumber\\
&\times&  \Big[
 [g^{(2)}_{F}(0) - 1] \langle a^\dagger a\rangle^2 
+ \frac{1}{4}[g^{(2)}_{D_1}(0) - 1]\langle b_1^\dagger b_1 \rangle^2 \nonumber\\&
 +& \frac{1}{4}[g^{(2)}_{D_2}(0) - 1] \langle b_2^\dagger b_2\rangle^2
+ \frac{1}{2} \langle b_1^\dagger b_1 b_2^\dagger b_2 \rangle \nonumber\\&
 -& \langle a^\dagger a b_1^\dagger b_1 \rangle
-  \langle a^\dagger a b_2^\dagger b_2 \rangle   
\Big].\label{kappa2}
\end{eqnarray}
See Appendix.~\ref{Appx-B-two_detector} for the details relevant for arbitrary detector preparations, which is lengthier, but potentially useful.  
We note that the coherence functions, $  g^{(2)}(0) = 1+\frac{Q}{\langle N\rangle} =\frac{\langle N(N-1)\rangle}{\langle N\rangle^2} $ appear. 
 Now assuming both detectors are at thermal state with an equal number of thermal quanta $n_{th}^{D}$, and applying small angle approximation, the cross correlation reduces to the form,
\begin{equation}
\kappa_2 = (\gamma_0\tau)^2
 \left\{\left[g^{(2)}_F(0)-1\right] \langle a^\dagger a \rangle^2
- 2 \langle a^\dagger a \rangle \, n_{th}^{D}
+ (n_{th}^{D})^2
\right\}.   \label{kappa2t}
\end{equation}
The leading contribution $ \kappa_2 = (\gamma_0 \,\tau)^2
 \left[g^{(2)}_F(0)-1\right] \langle a^\dagger a \rangle^2$ is of course what one would obtain if the detectors were to be initialized in their quantum ground states (see Ref.~\cite{Manikandan:2025lfx}). It allows to discriminate super-Poissonian states of the radiation field (examples include thermal states and squeezed vacuum states) from a coherent state. The important point we make in addition is that for Poissonian and super-Poissonian states of the radiation field for which $g^{2}(0)\geq 1$, if the detectors in the bolometric regime are weakly coupled ($\gamma_0\tau \ll 1$) but the incoming field intensity is such that the field is detectable with high chance ($\gamma_0\tau\langle a^\dagger  a\rangle\sim 1$), then it appears that the radiation field contribution to detector-cross-correlations dominate substantially over the inherent detector contributions, provided $\langle a^\dagger a\rangle \gg n_{th}^{D}$. This suggests that quantum ground state cooling of detectors is not necessary. 
 Our interest is also in the extreme limit relevant for graviton detectors, $\gamma_0\tau\rightarrow 0, \langle N\rangle \rightarrow \infty:\gamma_0\tau\langle N\rangle \rightarrow 1$, and we see that only the leading order term in $\kappa_2$  will survive at any finite $n_{th}^D$. Various other factors will require a small $n_{th}^D$, including for maintaining high-quality of the detector mode, their state of matter, etc., so our point (throughout the manuscript) is only that a small but finite $n_{th}^D$ appears to be tolerable, and absolute quantum ground state cooling may not be necessary.
 
Importantly, this statistical advantage is not present if the field couples strongly to the detectors, for example, the case for good single photon detectors ($\gamma_0\tau\sim 1$). In that regime, quantum ground state cooling of the detectors will be a must, since all terms in the cross correlation, Eq.~\eqref{kappa2}, would be of the same order in the functional detector regime, corresponding to a few quanta of the incoming radiation field owing to the high quantum-to-click conversion efficiency of strongly coupled detectors. 

 As an example, let us consider an incoming thermal state of a weak field with mean quanta $n^{F}_{\text{th}}$, using $\langle a^\dagger a\rangle = n^{F}_{\text{th}}$ and $\langle a^{\dagger 2} a^2\rangle = 2 (n^{F}_{\text{th}})^2$, $\kappa_2$ reduces to
\begin{equation}
 \kappa_2 = (\gamma_0 \tau)^2 (n^F_{th} - n_{th}^{D})^2.
 \end{equation}
 We see that the cross-correlator $\kappa_2$ vanishes only when $ n_{th}^{D} = n^F_{\text{th}}$, which can even be an unrealistic limit for weakly coupled detectors that respond only in the presence of a large number of incoming radiation field quanta $n^F_{th}\gg 1$. 
 In contrast, it is also evident that for good single photon detectors, the functional detector regime correspond to $\langle a^\dagger a\rangle \sim 1,$ and even a single thermal quanta in the detector, say a good bolometric detector~\cite{Sarkar:2025byz,Catto:2023ouc,Kokkoniemi:2020mou,Karimi:2024cog,Charaev:2024ptd} can have the same order effect in cross-correlation as the incoming field, suggesting that the strategy only work in the weakly coupled bolometric detection regime. 

 We also evaluate the detector-cross-correlations for phase-sensitive~\cite{wiseman_interpretation_1993,wiseman_quantum_1993,wiseman_quantum_1996,wiseman_quantum_2009,wiseman_stochastic_1993,lewalle_measuring_2020,caves_quantum-mechanical_1987} and phase-preserving~\cite{arthurs_simultaneous_1965} measurements, which are relevant to probe sub-Poissonian quantum statistics of the incoming radiation field using a weakly coupled detector in the bolometric regime. Note that for sub-Poissonian states such as Fock state $|n\rangle$ for which $g^{2}(0)=1-1/n$, the first term of Eq.~\eqref{kappa2t} will be smaller compared to the remaining terms, making it difficult to tell apart from a coherent state, using the correlation of counts, in the large $n$ limit. Phase measurements offer complementary means to overcome this~\cite{Manikandan:2025qgv,Manikandan:2025hlz}. For phase-sensitive quantum measurements, the correlators are given by (assuming units where the zero-point length of the detectors to be one. See Appendix.~\ref{App_Homodyne} for details),
 \begin{eqnarray}
&&\langle x_1 x_2 \rangle
-
\langle x_1\rangle\langle x_2\rangle
\approx
\frac{\gamma_0\tau}{2}\,
\bigg\{2\Big[\langle (\Delta P)^2\rangle -\frac{1}{2}\Big]\nonumber\\&&
-\Big[\langle (\Delta X_1)^2\rangle -\frac{1}{2}\Big]
-\Big[\langle (\Delta X_2)^2\rangle -\frac{1}{2}\Big]\bigg\}.
\end{eqnarray}
For thermal detectors, this reduces to $\langle x_1 x_2 \rangle
-
\langle x_1\rangle\langle x_2\rangle = \gamma_0\tau \Big[\langle (\Delta P)^2\rangle -\frac{1}{2}\Big]-\gamma_0\tau n_{th}^D.$  We see that similar conclusions hold, especially in the extreme limit $\gamma_0\tau\rightarrow 0, \langle N\rangle \rightarrow \infty:\gamma_0\tau\langle N\rangle \rightarrow 1$ of interest here, that any finite $n_{th}^D$ is tolerable. The results for phase-preserving measurements are presented in Appendix.~\ref{App_Heterodyne}. Extending these to the continuously monitored detector regime~\cite{Karmakar:2021aba} is also of great interest, and we defer it to future work.  

 The generalization to three or more detectors, although somewhat more involved, is straightforward. To demonstrate this, we consider three detectors prepared in identical thermal states of thermal quanta $n_{th}^{D}$. We can define the relevant third-order symmetric correlator as (see Appendix.~\ref{app_3D}),
 \begin{eqnarray}
\kappa_3 &\equiv& \langle (N_{1}-\langle N_1\rangle)(N_{2}-\langle N_2\rangle)(N_{3}-\langle N_3\rangle)\rangle\nonumber\\ 
&=& \langle N_1 N_2 N_3 \rangle 
- \langle N_1 \rangle \langle N_2 N_3 \rangle
- \langle N_2 \rangle \langle N_1 N_3 \rangle \nonumber \\
&-& \langle N_3 \rangle \langle N_1 N_2 \rangle
+ 2 \langle N_1 \rangle \langle N_2 \rangle \langle N_3 \rangle\nonumber\\
&=&\frac{\sin^6\left(\sqrt{3\gamma_0\tau}\right)}{27} \Big[
\langle a^{\dagger 3} a^3 \rangle 
- 3 \langle a^{\dagger 2} a^2 \rangle \big(\langle a^\dagger a \rangle + 2n_{th}^{D}\big)
\nonumber \\
& +& 2 \left\{ 
\langle a^\dagger a \rangle^3 
+ 3 \langle a^\dagger a \rangle^2 n_{th}^{D} 
+ 3 \langle a^\dagger a \rangle (n_{th}^{D})^2 
- (n_{th}^{D})^3 
\right\}
\Big]\nonumber\\
\end{eqnarray}
This can also be written in terms of the coherence functions $g^{(2)}_F(0)$ and $g^{(3)}_F(0)=\langle (a^\dagger)^3a^3\rangle/(\langle a^\dagger a\rangle^3)$ as,
\begin{eqnarray}
  \kappa_3 
&=& \frac{\sin^6\left(\sqrt{3\gamma_0\tau}\right)}{27} \Big[
\langle a^{\dagger}a \rangle^3(g^{(3)}_F(0) -3 g^{(2)}_F(0) +2) \nonumber\\
&-&6 \langle a^\dagger a \rangle^2 n_{th}^{D} (g^{(2)}_F(0) -1) + 6 \langle a^\dagger a \rangle (n_{th}^D)^2 -2 (n_{th}^D)^3
\Big] \nonumber\\&\approx&(\gamma_0\tau)^3 \Big[
\langle a^{\dagger}a \rangle^3(g^{(3)}_F(0) -3 g^{(2)}_F(0) +2) \nonumber\\
&-&6 \langle a^\dagger a \rangle^2 n_{th}^{D} (g^{(2)}_F(0) -1) + 6 \langle a^\dagger a \rangle (n_{th}^D)^2 -2 (n_{th}^D)^3
\Big].\nonumber\\  
\end{eqnarray}
The conclusions are similar, that the coherence properties of the radiation field, including the third order coherence functions, can, in principle, be estimated, provided the coupling is very weak such that $\gamma_0\tau n_{th}^{D}\ll \gamma_0\tau \langle a^\dagger a\rangle \sim 1$. In the limit of quantum-ground-state cooled detectors, $n_{th}^{D}\rightarrow 0$, and we see that the cross-correlator probes the coherence functions as,
\begin{eqnarray}
    \kappa_3 =(\gamma_0\tau)^3\langle a^{\dagger}a \rangle^3 \left[
g^{(3)}_F(0) -3 g^{(2)}_F(0) +2 \right].
\end{eqnarray}
The leading order contribution to the correlator vanishes for a coherent state, and for strongly interacting fields for which $\gamma_0\tau\sim O(1)$, we would indeed require ground state cooling to infer the coherence functions using this method. This reaffirms our observation that, for weakly interacting fields, the requirement is not ground state cooling, rather only that $n_{th}^{D}\ll \langle a^\dagger a\rangle$, which is easily satisfied in the functional detector regime for gravitons, owing to their extremely weak coupling. If the radiation field is also prepared in a thermal state, we can approximate $\kappa_3$ as,
\begin{eqnarray}
    \kappa_3 =2(\gamma_0\tau)^3(n^{F}_{th}-n^{D}_{th})^3.
\end{eqnarray}
We again see that the cross-correlator vanishes identically when $n_{th}^{F} =n_{th}^{D}.$ However, for weakly coupled radiation fields (for which $\gamma_0\tau\ll 1$), detectors could produce a response only when $n_{th}^{F}\gg 1$. Hence, we would very likely be in the regime $n_{th}^{F} \gg n_{th}^{D}$, warranting the feasibility of the statistical tests proposed in this work, provided the detectors are of high quality.

Before we conclude, we again emphasize that there are many obvious physical motivations for requiring detectors maintained at low temperature, including to maintain their high quality, the phase of matter, and even for the practicalities of the cross-correlated quantum sensing we proposed. Our point is to make note of the added advantages at the barely functional bolometric detector regime, which suggest that absolute ground state cooling may not be necessary for the detectors of weak fields to probe the quantum noise characteristics of the indivisible quanta of the weak field, such as gravitons, using the symmetric correlators that we proposed.  Such correlated quantum sensing strategies may also have implications for entanglement-based tests that aim to probe quantum gravity in a different spin sector~\cite{Marletto:2017kzi,Bose:2017nin,Fragkos:2022tbm}, which is worth exploring further.

\noindent\textit{Conclusions.---} We have shown that independent detector cross-correlations between two or more high-quality, but barely functional resonant harmonic detectors operating in the bolometric regime offer a quantum statistical advantage for probing the complementary quantum noise characteristics of quantized radiation fields in the seemingly classical limit of large $\langle N\rangle$.  Given that probing the quantum properties of weakly interacting fields is a challenging task, the correlated quantum sensing strategies we proposed are of particular interest for testing the quantum character of weak fields that are yet to be detected, or fields that have been detected classically, but their quantum properties are yet to be verified, such as the gravitational radiation field using high-quality resonant mass-quadrupole oscillators as detectors. The optimal mass for the resonant mass detector at frequency $\omega$ for a given source with chirp mass $M_c$ is, $M=\frac{24{\pi }^{2}}{5}\frac{\hbar }{{h}_{0}^{2}{v}_{{s}}^{2}}{\left(\frac{G{M}_{{c}}}{2{c}^{3}}\right)}^{5/3}{\omega }^{8/3}\approx 15\text{kg}$~\cite{Tobar:2023ksi},
for  a Beryllium detector at 100Hz frequency, assuming the gravitational wave strain amplitude, $h_0\approx 2\times 10^{-22}$, $G$ is the Newton's constant, $c$ is the speed of light, and $v_s$ to be the speed of sound in Beryllium. The quantum ground state cooling of acoustic modes of such a detector can indeed be a challenge, which makes it relevant to have statistical probes of the type we proposed that do not require quantum ground state cooling, and only demand high quality factors. Broadly, our findings suggest that some of the challenging aspects for probing the indivisible quanta of weakly coupled radiation fields, including the gravitational radiation field, may be alleviated by the very fact that good detectors are fundamentally impossible for these quanta. 

\noindent\textit{Acknowledgements.---}
The authors acknowledge the support from the Department of Atomic Energy, Government of India, under Project Identification No. RTI4007. 

\noindent\textit{Author Contributions.---}
The work was conceptualized by SKM. Both KPA and SKM contributed to the calculations, analyzing the results, and writing the manuscript.
\appendix
\begin{widetext}
    \section{The Dynamics for Generic Field and Detector States}
    \label{Appx-A}
Here we evaluate the time evolution of the detector and the field, using the Glauber-Sudarshan $P$ representation. The initial state of field and detector is considered to be initially uncorrelated, given by, 
\begin{equation}
\rho(0) = \rho_{F}(0) \otimes \rho_{D}(0),
\end{equation}
where the state of the field and the detector can be expressed as,
\begin{equation}
    \rho_{i}(0) = \int d^2\alpha P_i(\alpha)|\alpha\rangle\langle\alpha|,~~i\in \{F,D\}.
\end{equation}
Hence,
\begin{equation}
\rho(0) 
= \int d^2\alpha \, d^2\beta \;
P_{F}(\alpha)\, P_{D}(\beta)\;
\ket{\alpha}\bra{\alpha} \otimes \ket{\beta}\bra{\beta}.
\end{equation}
The state of the field and detector is modified by the interaction unitary,
\begin{equation}
U_{I} 
= \exp\left[-i \theta\,
\left(a^\dagger b + b^\dagger a\right)\right], \qquad \text{where},\,
\theta=\sqrt{\gamma_0\tau}.
\label{1d_unitrary}
\end{equation}
After the interaction, the joint state of the field and detector is modified to $\rho' 
= U_{I} \rho(0) U_{I}^\dagger$.
To evaluate, it is sufficient to determine the action of $U_I$ on the state $\ket{\alpha}\ket{\beta}$.
Using,
\begin{align}
\ket{\alpha} &= e^{-\frac{|\alpha|^2}{2}} e^{\alpha a^\dagger}\ket{0},\\
\ket{\beta}  &= e^{-\frac{|\beta|^2}{2}} e^{\beta b^\dagger}\ket{0},
\end{align}
we write,
\begin{align}
U_{I} \ket{\alpha}\otimes\ket{\beta}
&= e^{-\frac{|\alpha|^2}{2}} e^{-\frac{|\beta|^2}{2}}
\, U_{I} e^{\alpha a^\dagger} e^{\beta b^\dagger}\ket{0}\otimes\ket{0}.
\label{1d_eff_trans}
\end{align}
We can now move $U_{I}$ through the exponential functions,
\begin{align}
U_{I} e^{\alpha a^\dagger} U_{I}^\dagger
&= U_{I} \sum_{n=0}^\infty \frac{(\alpha a^\dagger)^n}{n!} U_I^\dagger
= \sum_{n=0}^\infty \frac{\alpha^n}{n!}
\left(U_{I} a^\dagger U_{I}^\dagger\right)^n
= \exp\!\left[\alpha\, U_{I} a^\dagger U_{I}^\dagger\right],
\label{1d_full_tras}
\end{align}
where we use,
\begin{align}
U_{I} a^\dagger U_{I}^\dagger &= a^\dagger \cos\theta - i b^\dagger \sin\theta,
\label{1d_t1}
\\
U_{I} b^\dagger U_{I}^\dagger &= b^\dagger \cos\theta - i a^\dagger \sin\theta.
\label{1d_t2}
\end{align}
Substituting Eqs.~\eqref{1d_t1} and \eqref{1d_t2} into Eq.~\eqref{1d_eff_trans} and collecting terms yields,
\begin{align}
U_{I} \ket{\alpha}\otimes\ket{\beta}
&= e^{-\frac{|\alpha|^2}{2}} e^{-\frac{|\beta|^2}{2}}
\exp\!\left[\alpha(a^\dagger\cos\theta - i b^\dagger\sin\theta)\right]
\nonumber\\
&\quad\times
\exp\!\left[\beta(b^\dagger\cos\theta - i a^\dagger\sin\theta)\right]
\ket{0}\otimes\ket{0}\nonumber\\ 
&= e^{-\frac{|\alpha|^2}{2}} e^{-\frac{|\beta|^2}{2}}
\exp\!\left[(\alpha\cos\theta - i\beta\sin\theta)a^\dagger\right]
\nonumber\\
&\quad\times
\exp\!\left[(\beta\cos\theta - i\alpha\sin\theta)b^\dagger\right]
\ket{0}\otimes\ket{0}.
\end{align}
Using the coherent-state identity, we obtain,
\begin{align}
e^{(\alpha\cos\theta - i\beta\sin\theta)a^\dagger}\ket{0}
&= e^{\frac{1}{2}|\alpha\cos\theta - i\beta\sin\theta|^2}
\ket{\alpha\cos\theta - i\beta\sin\theta},\\
e^{(\beta\cos\theta - i\alpha\sin\theta)b^\dagger}\ket{0}
&= e^{\frac{1}{2}|\beta\cos\theta - i\alpha\sin\theta|^2}
\ket{\beta\cos\theta - i\alpha\sin\theta}.
\end{align}
Therefore,
\begin{equation}
U_{I}\ket{\alpha}_a\ket{\beta}_b
= \ket{\alpha'}_a \otimes \ket{\beta'}_b,
\end{equation}
where we can identify,
\begin{align}
\alpha' &= \alpha\cos\theta - i\beta\sin\theta,\\
\beta'  &= \beta\cos\theta - i\alpha\sin\theta.
\end{align}
This transformation (which also shows how a beam-splitter mixes its two inputs~\cite{Gerry_Knight_2004}) describes how a general state of field and a single detector expressed in the Glauber-Sudarshan $P$ representation evolves under the resonant harmonic interaction unitary given in Eq.~\eqref{1d_unitrary} through its action on a product of arbitrary coherent states.
The time-evolved state of the field and detector after the interaction can then be written as, 
\begin{equation}
\rho' 
= \int d^2\alpha \, d^2\beta \;
P_{F}(\alpha)\, P_{D}(\beta)\;
 \ket{\alpha'}\bra{\alpha'} \otimes \ket{\beta'}\bra{\beta'}.
\end{equation}
Now, the probability $P_n$ that the detector registers $n$ quanta is given by projecting the detector onto the number (Fock) basis $\{\ket{n}\}$ and tracing over the field, given by,

\begin{eqnarray}
P_{n} &=& \int d^2\alpha\, d^2\beta P_{F}(\alpha)\, P_{D}(\beta)
\frac{|\beta'|^{2n}}{n!} e^{-|\beta'|^2}\nonumber\\&=&
\frac{1}{n!}
\int d^2\alpha d^2\beta P_{F}(\alpha) P_{D}(\beta)\nonumber\\
&\times&|\beta\cos\theta- i\alpha\sin\theta|^{2n}
e^{-|\beta\cos\theta - i\alpha\sin\theta|^2}.\label{EqStat1}
\end{eqnarray}
where $\theta = \sqrt{\gamma_0 \tau}$. 
This double-convolution of the Poisson distribution weighted by the $P$ functions describing the quantum states of the radiation field and the detector represents the general counting probability for an arbitrary quantum state of both the radiation field and the detector. From Eq.~\eqref{EqStat1}, we can compute the mean number of detected counts, $\bar{n} = \langle b^\dagger b \rangle$, given by,
\begin{equation}
\bar{n} = \sum_{n=0}^\infty n P_{n} = \int d^2\alpha d^2\beta P_{F}(\alpha) P_{D}(\beta)|\beta'|^2.
\end{equation}
Upon simplification, the expression for the mean number of quanta reduces to the form,
\begin{equation}
\bar{n}
= c^2 \langle b^\dagger b\rangle
+ s^2 \langle a^\dagger a\rangle
+ i cs \left(\langle b a^\dagger\rangle
- \langle a b^\dagger\rangle\right).
\end{equation}
where $ c = \cos(\sqrt{\gamma_0 \tau}),$ and $ s=\sin(\sqrt{\gamma_0 \tau}).$

The number variance is calculated as given below,
\begin{equation}
(\Delta n)^2
= \langle n(n-1)\rangle
+ \langle n\rangle
- \langle n\rangle^2.
\end{equation}
where,
\begin{equation}
\langle n(n-1)\rangle=\int d^2\alpha d^2\beta P_{F}(\alpha) P_{D}(\beta)|\beta'|^4.
\end{equation}
The variance can be written as,
\begin{equation}
(\Delta n)^2
= \langle |\beta'|^4\rangle
+ \langle |\beta'|^2\rangle
- \langle |\beta'|^2\rangle^2.
\end{equation}
Here the averages $\langle\cdot\rangle$ denote integration over the $P$-functions of the field and detector, i.e.,
\[
\langle f(\alpha,\beta)\rangle
= \int d^2\alpha\, d^2\beta
P_F(\alpha)P_D(\beta)
f(\alpha,\beta).
\]

We will focus on the case where the detector is initially in a zero-mean-quadrature state, such as a thermal state. 
For such a detector, the cross terms such as \((\beta\alpha^*-\alpha\beta^*)\) will vanish on averaging. 
This yields the following average number of quanta in the detector,
\[
\bar n
= \cos^2(\sqrt{\gamma_0 \tau})\langle N_b\rangle
+\sin^2(\sqrt{\gamma_0 \tau})\langle N_a\rangle.
\]
with $\langle N_b\rangle = \langle b^\dagger b\rangle$ and $\langle N_a\rangle = \langle a^\dagger a\rangle$. Similarly, the observed quantum noise or variance of counts in the detector is given by (after systematically evaluating the average over zero-mean states),
\begin{equation}
\begin{split}
(\Delta n)^2
&= \cos^2(\sqrt{\gamma_0 \tau})\,\langle N_b\rangle
+ \sin^2(\sqrt{\gamma_0 \tau})\,\langle N_a\rangle \\
&\quad + \cos^4(\sqrt{\gamma_0 \tau})\,\langle N_b\rangle Q_b
+ \sin^4(\sqrt{\gamma_0 \tau})\,\langle N_a\rangle Q_a \\
&\quad + 2 \cos^2(\sqrt{\gamma_0 \tau}) \sin^2(\sqrt{\gamma_0 \tau}) \langle N_a \rangle \langle N_b \rangle.
\end{split}
\end{equation}
\section{Two Detector Model}
\label{Appx-B-two_detector}
The interaction Hamiltonian for the radiation field and two detectors simultaneously coupled to the radiation field is given by,
\begin{equation}
H_I \tau
=
\hbar \sqrt{\gamma_0 \tau}
\sum_{j=1}^{2}
\left(
a^\dagger b_j
+
b_j^\dagger a
\right).
\label{two_det}
\end{equation}
As in Appendix~\ref{Appx-A}, we assume that the field and detector states are initially uncorrelated. The total initial state can therefore be written as,
\begin{equation}
\rho(0)
=
\rho_F
\otimes
\rho_{D_1}
\otimes
\rho_{D_2}.
\label{2d_ini_rho}
\end{equation}
Using the Glauber-Sudarshan $P$ representation introduced in Appendix~\ref{Appx-A}, this state becomes,
\begin{align}
\rho(0)
=
\int d^2\alpha\,
d^2\beta_1\,
d^2\beta_2\;
&
P_F(\alpha)
P_1(\beta_1)
P_2(\beta_2)
\nonumber\\
\times\;
&
\ket{\alpha}\bra{\alpha}
\otimes
\ket{\beta_1}\bra{\beta_1}
\otimes
\ket{\beta_2}\bra{\beta_2}.
\label{2d_in_rho_P}
\end{align}
To evaluate the action of the unitary corresponding to the interaction Hamiltonian in Eq.~\eqref{two_det} on the state given in Eq.~\eqref{2d_in_rho_P}, we first 
Introduce the collective modes $d_{+}$ and $d_{-}$, defined by,
\begin{align}
  d_{+} &= \frac{1}{\sqrt{2}}\left(b_1 + b_2\right): \left[d_+, d_+^\dagger\right] =1,\\
  d_{-} &= \frac{1}{\sqrt{2}}\left(b_1 - b_2\right): \left[d_-, d_-^\dagger\right] =1.
\end{align}
The corresponding interaction Hamiltonian reduces to, 
\begin{equation}
H_I \tau
=
\hbar \sqrt{2\gamma_0 \tau}
\left(
a^\dagger d_+
+
d_+^\dagger a
\right).
\label{HI2_dominant}
\end{equation}
Thus, we see that only the symmetric (bright) mode couples to the field, while the antisymmetric mode remains unchanged. Therefore, we calculate the action of the interaction unitary corresponding to the Hamiltonian in Eq.~\eqref{HI2_dominant} on an initial state,
\begin{equation}
   \ket{\psi(0)} = \ket{\alpha}\ket{\beta_1}\ket{\beta_2},
\end{equation}
where we define,
\begin{align}
\beta_+
&=
\frac{\beta_1+\beta_2}{\sqrt{2}},
\\
\beta_-
&=
\frac{\beta_1-\beta_2}{\sqrt{2}}.
\end{align}
Using the single-mode result derived in Appendix~\ref{Appx-A},
\begin{equation}
U_I
\ket{\alpha}
\ket{\beta_+}
\ket{\beta_-}
=
\ket{\alpha'}
\ket{\beta_+'}
\ket{\beta_-},
\label{d2_basis_trans}
\end{equation}
where we can identify,
\begin{align}
\alpha'
&=
\alpha \cos(\sqrt{2\gamma_0\tau})
-
i\beta_+ \sin(\sqrt{2\gamma_0\tau}),
\\
\beta_+'
&=
\beta_+ \cos(\sqrt{2\gamma_0\tau})
-
i\alpha \sin(\sqrt{2\gamma_0\tau}).
\end{align}
Inverting these equations back to original modes gives,
\begin{align}
\alpha'&=\alpha\cos(\sqrt{2\gamma_0\tau}) -\frac{i}{\sqrt{2}}(\beta_{1} + \beta_{2})\sin(\sqrt{2\gamma_0\tau}),\\
\beta_{1}'&=\frac{(1+\cos(\sqrt{2\gamma_0\tau})}{2}\beta_{1}-
\frac{(1-\cos(\sqrt{2\gamma_0\tau})}{2}\,\beta_2-
\frac{i}{\sqrt{2}}\alpha\sin(\sqrt{2\gamma_0\tau}),\\
\beta_{2}'&=\frac{(1+\cos(\sqrt{2\gamma_0\tau})}{2}\beta_2 -\frac{(1-\cos(\sqrt{2\gamma_0\tau}))}{2}\beta_1-\frac{i}{\sqrt{2}}\alpha\sin(\sqrt{2\gamma_0\tau}).
\end{align}
Using trigonometric identity,
\begin{align}
\alpha'&=\alpha\cos(\sqrt{2\gamma_0\tau})-\frac{i}{\sqrt{2}}(\beta_{1} + \beta_{2})\sin(\sqrt{2\gamma_0\tau}),\\
\beta_{1}'&=\beta_{1}\cos^2(\frac{\sqrt{2\gamma_0\tau}}{2})-
\beta_2\sin^2(\frac{\sqrt{2\gamma_0\tau}}{2})-\frac{i}{\sqrt{2}}\alpha \sin(\sqrt{2\gamma_0\tau}),\label{two_det_trans_1}\\
\beta_{2}'&=\beta_{2}\cos^2(\frac{\sqrt{2\gamma_0\tau}}{2})-
\beta_{1}\sin^2(\frac{\sqrt{2\gamma_0\tau}}{2})-
\frac{i}{\sqrt{2}}\alpha\sin(\sqrt{2\gamma_0\tau}).
\label{two_det_trans_2}
\end{align}
These transformations form the key relations used in the calculations of moments and correlations for the two-detector model discussed in the main text.

\subsubsection{Joint Counting Statistics}
To calculate the correlation or cross-correlation between the counts of two detectors, we first need to evaluate the joint counting probability. For a coherent detector output states $\ket{\beta_i'}$, the statistics is Poissonian. 
\begin{equation}
P(n_i|\beta_i')
=\frac{|\beta_i'|^{2n_i}}{n_i!}e^{-|\beta_i'|^2}.
\end{equation}
The joint probability of observing $n_1$ quanta in detector $1$ and $n_2$ quanta in detector $2$  is therefore given by,
\begin{align}
P(n_1,n_2)
&=\int d^2\alpha d^2\beta_1 d^2\beta_2
P_{F}(\alpha)P_{1}(\beta_1)P_{2}(\beta_2)
\frac{|\beta_1'|^{2n_1}}{n_1!}e^{-|\beta_1'|^2}
\frac{|\beta_2'|^{2n_2}}{n_2!}e^{-|\beta_2'|^2}.
\end{align}
The mean detector counts are calculated using,
\begin{equation}
\langle N_i \rangle
=
\sum_{n_1,n_2}
n_i
P(n_1,n_2)
=
\int d^2\alpha\, d^2\beta_1\, d^2\beta_2\;
P_F P_1(\beta_1) P_2(\beta_2)
|\beta_i'|^2.
\label{eq:Ni_mean}
\end{equation}
Similarly,
\begin{align}
\langle N_{1} N_{2} \rangle
= \langle b_1^\dagger b_1 b_2^\dagger b_2 \rangle
= \int d^2\alpha d^2\beta_1 d^2\beta_2
P_F(\alpha) P_1(\beta_1) P_2(\beta_2)|\beta_1'|^2 |\beta_2'|^2,
\end{align}
where assuming $\theta_0 =\sqrt{2 \gamma_0 \tau}$,
\begin{align}
|\beta_1'|^2=
&|\beta_1|^2 \cos^4\left(\frac{\theta_0}{2}\right)
- \beta_1^* \beta_2 \cos^2\left(\frac{\theta_0}{2}\right)\sin^2\left(\frac{\theta_0}{2}\right)
- \beta_1 \beta_2^* \cos^2\left(\frac{\theta_0}{2}\right)\sin^2\left(\frac{\theta_0}{2}\right) \\ \nonumber
&+ |\beta_2|^2 \sin^4\left(\frac{\theta_0}{2}\right)
+ \frac{i \alpha^* \beta_1 \cos^2\left(\frac{\theta_0}{2}\right)\sin\theta_0}{\sqrt{2}}
- \frac{i \alpha \beta_1^* \cos^2\left(\frac{\theta_0}{2}\right)\sin\theta_0}{\sqrt{2}} \\ \nonumber
&- \frac{i \alpha^* \beta_2 \sin^2\left(\frac{\theta_0}{2}\right)\sin\theta_0}{\sqrt{2}}
+ \frac{i \alpha \beta_2^* \sin^2\left(\frac{\theta_0}{2}\right)\sin\theta_0}{\sqrt{2}}
+ \frac{1}{2} |\alpha|^2 \sin^2\theta_0,
\end{align}
and,
\begin{align}
|\beta_2'|^2=&|\beta_2|^2 \cos^4\left(\frac{\theta_0}{2}\right)
- \beta_1^* \beta_2 \cos^2\left(\frac{\theta_0}{2}\right)\sin^2\left(\frac{\theta_0}{2}\right)
- \beta_1 \beta_2^* \cos^2\left(\frac{\theta_0}{2}\right)\sin^2\left(\frac{\theta}{2}\right) \\\nonumber
&+ |\beta_1|^2 \sin^4\left(\frac{\theta_0}{2}\right)
+ \frac{i \alpha^* \beta_2 \cos^2\left(\frac{\theta_0}{2}\right)\sin\theta_0}{\sqrt{2}}
- \frac{i \alpha \beta_2^* \cos^2\left(\frac{\theta_0}{2}\right)\sin\theta_0}{\sqrt{2}} \\ \nonumber
&- \frac{i \alpha^* \beta_1 \sin^2\left(\frac{\theta_0}{2}\right)\sin\theta_0}{\sqrt{2}}
+ \frac{i\alpha \beta_1^* \sin^2\left(\frac{\theta_0}{2}\right)\sin\theta_0}{\sqrt{2}}
+ \frac{1}{2} |\alpha|^2 \sin^2\theta_0.
\end{align}
Now to calculate the correlated counts, we need to evaluate $\langle |\beta'_1|^2|\beta'_2|^2 \rangle $.
The product $|\beta'_1|^2|\beta'_2|^2$ contains many terms, and the expectation value can be calculated by evaluating the integrals over the $P$ functions of the field and two detectors over these terms. In this work, we restrict ourselves to a zero-mean state of detectors, such as a thermal state, which allows several terms to vanish and thereby simplify the calculation. However, for a more general choice of field and detector state, all terms, in principle, must be retained. 
We now calculate the detector cross correlation using,
$\langle N_1 N_2 \rangle - \langle N_1 \rangle \langle N_2 \rangle $ = $\langle |\beta'_1|^2|\beta'_2|^2  \rangle - \langle |\beta'_1|^2 \rangle \langle |\beta'_2|^2  \rangle $ . Here, assuming
detectors are initially prepared in independent zero-mean states, such as a thermal state, simplifying the cross-correlation to,
\begin{align}
\langle N_1 N_2 \rangle - \langle N_1 \rangle \langle N_2 \rangle
&= \frac{1}{4} \sin^4\theta_0 
\left( \langle |\alpha|^4 \rangle -\langle |\alpha|^2 \rangle^2 \right)\nonumber\\ 
&\quad + \cos^4\frac{\theta_0}{2}\, \sin^4\frac{\theta_0}{2}
\left( \langle |\beta_1|^4 \rangle - \langle |\beta_1|^2 \rangle^2 \right) \nonumber\\
&\quad + \sin^4\frac{\theta_0}{2}\, \cos^4\frac{\theta_0}{2}
\left( \langle |\beta_2|^4 \rangle - \langle |\beta_2|^2 \rangle^2 \right)\nonumber\\
&\quad + 2 \cos^4\frac{\theta_0}{2}\, \sin^4\frac{\theta_0}{2} \langle |\beta_1|^2  |\beta_2|^2 \rangle \nonumber\\
&\quad - \cos^2\frac{\theta_0}{2}\, \sin^2\frac{\theta_0}{2}\, \sin^2\theta_0  \langle |\alpha|^2  |\beta_1|^2 \rangle \nonumber\\
&\quad - \cos^2\frac{\theta_0}{2}\, \sin^2\frac{\theta_0}{2}\, \sin^2\theta_0  \langle |\alpha|^2  |\beta_2|^2 \rangle.
\end{align}
Using small-angle approximation, we further simplify the above expression for cross correlation to,
\begin{align}
\langle N_1 N_2 \rangle - \langle N_1 \rangle \langle N_2 \rangle&\approx \frac{\theta_0^4}{4}
\left( \langle |\alpha|^4 \rangle - \langle |\alpha|^2 \rangle^2 \right)\nonumber \\
&\quad + \frac{\theta_0^4}{16}
\left( \langle |\beta_1|^4 \rangle - \langle |\beta_1|^2 \rangle^2 \right) \nonumber\\
&\quad + \frac{\theta_0^4}{16}
\left( \langle |\beta_2|^4 \rangle - \langle |\beta_2|^2 \rangle^2 \right)\nonumber\\
&\quad + \frac{\theta_0^4}{8} \langle |\beta_1|^2 |\beta_2|^2 \rangle - \frac{\theta_0^4}{4} \langle |\alpha|^2 |\beta_1|^2 \rangle - \frac{\theta_0^4}{4} \langle |\alpha|^2 |\beta_2|^2 \rangle.
\end{align}
Finally using $\theta_0 =\sqrt{2 \gamma_0 \tau}$ ,
\begin{align}
\langle N_1 N_2 \rangle - \langle N_1 \rangle \langle N_2 \rangle
&= \gamma_0^2 \tau^2
\left( \langle |\alpha|^4 \rangle - \langle |\alpha|^2 \rangle^2 \right) \nonumber \\
&\quad + \frac{\gamma_0^2 \tau^2}{4}
\left( \langle |\beta_1|^4 \rangle - \langle |\beta_1|^2 \rangle^2 \right) \nonumber \\
&\quad + \frac{\gamma_0^2 \tau^2}{4}
\left( \langle |\beta_2|^4 \rangle - \langle |\beta_2|^2 \rangle^2 \right)\nonumber\\
&\quad + \frac{\gamma_0^2 \tau^2}{2} \langle |\beta_1|^2 |\beta_2|^2 \rangle - \gamma_0^2 \tau^2 \langle |\alpha|^2 |\beta_1|^2 \rangle - \gamma_0^2 \tau^2 \langle |\alpha|^2 |\beta_2|^2 \rangle. 
\end{align}
To small $\gamma_0\tau$ this agrees with what we have in the main text.
\section{Three-Detector Model\label{app_3D}}
The interaction Hamiltonian for the radiation field interacting simultaneously with three detectors can be written as,
\begin{equation}
H_I \tau
=
\hbar \sqrt{\gamma_0 \tau}
\sum_{j=1}^{3}
\left(
a^\dagger b_j
+
b_j^\dagger a
\right).
\end{equation}
A straightforward extension of single and two detector cases allows us to write the uncorrelated initial state of the radiation field and three detectors as, 
\begin{equation}
    \rho(0) = \rho_F \otimes \rho_{D_1}\otimes\rho_{D_2}\otimes\rho_{D_3}.
\end{equation}
Now, as we did in the two-detector case, we define normal modes for this three-detector case as follows:
\begin{align}
d_+ &= \frac{1}{\sqrt{3}} (b_1 + b_2 + b_3), \\
d_1 &= \frac{1}{\sqrt{2}} (b_1 - b_2), \\
d_2 &= \frac{1}{\sqrt{6}} (b_1 + b_2 - 2b_3).
\end{align}
This choice of new modes satisfies the canonical bosonic commutation relations, 
\begin{equation}
[d_i, d_j^\dagger] = \delta_{ij}.
\end{equation}
In terms of these collective modes, the interaction Hamiltonian for the three-detector model reduces to,
\begin{equation}
H_{I}\tau = \hbar \sqrt{3 \gamma_0 \tau} \left(
a d_+^\dagger + d_+ a^\dagger
\right).
\label{3d_noraml}
\end{equation}
So the action of the interaction unitary corresponding to the Hamiltonian in Eq.~\eqref{3d_noraml}  on the initial state $\ket{\alpha}\ket{\beta_1}\ket{\beta_2}\ket{\beta_3}$, leading to,
\begin{equation}
    U_I \ket{\alpha}\ket{\beta_1}\ket{\beta_2}\ket{\beta_3}= U_I \ket{\alpha}\ket{\beta_+}\ket{\beta_1^{-}}\ket{\beta_2^{-}},
\end{equation}
and we denote, $ U_I \ket{\alpha}\ket{\beta_+}\ket{\beta_1^{-}}\ket{\beta_2^{-}} = \ket{\alpha^{'}}\ket{\beta_+^{'}}\ket{\beta_1^{-'}}\ket{\beta_{2}^{-'}}$.
Here as well, the interaction unitary contains only terms involving mode $a$ and the bright mode represented by $d_{+}$. Therefore the states $\ket{\beta_1^{-'}}$ and $\ket{\beta_2^{-'}}$ remains unchanged under the action of the unitary. 
\begin{align}
\alpha' &= \alpha \cos \phi - i \beta_+ \sin \phi, \\
\beta'_+ &= \beta_+ \cos \phi - i \alpha \sin \phi, \\
\beta_1^{-'} &= \beta_1^{-},\\
\beta_2^{-'} &= \beta_2^{-}.\\
\end{align}
where $\phi = \sqrt{3\gamma_0\tau} $ for the three detector case. 
\begin{align}
 \beta_+ &= \frac{1}{\sqrt{3}}(\beta_1+\beta_2+\beta_3),\\
\beta_1^{-} &= \frac{1}{\sqrt{2}}(\beta_1-\beta_2),\\
\beta_2^{-} &= \frac{1}{\sqrt{6}}(\beta_1 + \beta_2 - 2 \beta_3).
\end{align}
By inverting the above equations to the original mode, we obtain the general state result for the three-detector case:
\begin{align}
\alpha' &= \alpha \cos\phi 
- \frac{i}{\sqrt{3}} (\beta_1 + \beta_2 + \beta_3)\sin\phi, \\
\beta_1' &= \frac{2 + \cos\phi}{3}\beta_1 
- \frac{1 - \cos\phi}{3}\,\beta_2 
- \frac{1 - \cos\phi}{3}\beta_3 
- \frac{i \sin\phi}{\sqrt{3}}\,\alpha, \\
\beta_2' &= -\frac{1 - \cos\phi}{3}\beta_1 
+ \frac{2 + \cos\phi}{3}\beta_2 
- \frac{1 - \cos\phi}{3}\beta_3 
- \frac{i \sin\phi}{\sqrt{3}}\alpha, \\
\beta_3' &= -\frac{1 - \cos\phi}{3}\beta_1 
- \frac{1 - \cos\phi}{3}\beta_2 
+ \frac{2 + \cos\phi}{3}\beta_3 
- \frac{i \sin\phi}{\sqrt{3}}\alpha.
\end{align}
The above relations are used throughout the subsequent calculations of counting statistics and correlations in the three-detector model.
\subsubsection{Joint Counting Statistics}
For three detector case joint probability to get $n_1$ click on detector 1, $n_2$ click on detector 2  and $ n_3$ clicks on detector 3 is given by 
\begin{align}
P(n_1,n_2,n_3)
&=\int d^2\alpha d^2\beta_1 d^2\beta_2d^2\beta_3
P_{F}(\alpha)P_{1}(\beta_1)P_{2}(\beta_2)P_{3}(\beta_3)\nonumber\\
&\quad\times
\frac{|\beta_1'|^{2n_1}}{n_1!}e^{-|\beta_1'|^2}
\frac{|\beta_2'|^{2n_2}}{n_2!}e^{-|\beta_2'|^2}
\frac{|\beta_3'|^{2n_3}}{n_3!}e^{-|\beta_3'|^2}.
\end{align}
Using Glauber Sudarshan $P$ representation, the mean detector counts are calculated by
\begin{align}
\langle N_i \rangle
= \langle b_i^\dagger b_i \rangle
= \int d^2\alpha d^2\beta_1 d^2\beta_2 d^2\beta_3
P_F(\alpha) P_1(\beta_1)P_2(\beta_2)P_3(\beta_3)|\beta_i'|^2,
\end{align}
where $P_F(\alpha)$, $P_1(\beta_1)$, $P_2(\beta_2)$ and $P_3(\beta_3)$ are the $P$ functions of the field, detector $1$, detector $2$ and detector $3$, respectively.
To compute correlations, we also require joint moments:
\begin{align}
\langle N_i N_j \rangle 
= \langle b_i^\dagger b_i \, b_j^\dagger b_j \rangle.
\end{align}
In the $P$-representation, this becomes
\begin{align}
\langle N_i N_j \rangle
= \int d^2\alpha \prod_{k=1}^3 d^2\beta_k 
P_F(\alpha)\prod_{k=1}^3 P_k(\beta_k)
|\beta_i'|^2 |\beta_j'|^2.
\end{align}
\subsubsection{Third-Order Correlation}
To characterize the correlation between detectors, we need to calculate the symmetric third-order correlation for the three-detector case. A third-order moment is given by
\begin{align}
\langle N_1 N_2 N_3 \rangle
= \langle b_1^\dagger b_1 \, b_2^\dagger b_2 \, b_3^\dagger b_3 \rangle.
\end{align}
Using the $P$-representation, this maybe evaluated as
\begin{align}
\langle N_1 N_2 N_3 \rangle
= \int d^2\alpha \prod_{k=1}^3 d^2\beta_k
P_F(\alpha)\prod_{k=1}^3 P_k(\beta_k)
|\beta_1'|^2 |\beta_2'|^2 |\beta_3'|^2.
\end{align}
Since the third order moments contain contributions from lower order correlation, we introduce the following symmetric third order correlation function to isolate the genuine three detector correlations. The symmetric third-order correlation function (cumulant) is defined as,
\begin{align}
\kappa_3 
&= \langle N_1 N_2 N_3 \rangle 
- \langle N_1 \rangle \langle N_2 N_3 \rangle
- \langle N_2 \rangle \langle N_1 N_3 \rangle \nonumber \\
&\quad - \langle N_3 \rangle \langle N_1 N_2 \rangle
+ 2 \langle N_1 \rangle \langle N_2 \rangle \langle N_3 \rangle.
\end{align}
This average involves integration over many terms for generic detector preparations, even for thermal initial conditions of the detector. The corresponding correlator and statistical inferences for thermal initial states of the detector are discussed in the main text.

\section{Correlations in Homodyne Measurement\label{App_Homodyne}}
In this section, we derive the correlation between the homodyne measurement outcomes of the two-detector model. As in the case of number detection, the state of the field and the two detectors are initially assumed to be uncorrelated so that the total density matrix can be written as a product state. In the $P$ representation, this takes the form,
\begin{align}
\rho 
= \int d^2\alpha \, d^2\beta_1 \, d^2\beta_2 
P(\alpha)\, P(\beta_1)\, P(\beta_2)
|\alpha\rangle \langle \alpha| \otimes |\beta_1\rangle \langle \beta_1| \otimes |\beta_2\rangle \langle \beta_2|.
\end{align}
After interaction, each coherent state amplitude is transformed to $\alpha'$, $\beta_1'$, and $\beta_2'$, as discussed earlier in the main text and Appendix.~\ref{Appx-B-two_detector}. The corresponding transformed state is given by,
\begin{align}
\rho' 
= \int d^2\alpha \, d^2\beta_1 \, d^2\beta_2 
P(\alpha)\, P(\beta_1)\, P(\beta_2)
|\alpha'\rangle \langle \alpha'| \otimes |\beta_1'\rangle \langle \beta_1'| \otimes |\beta_2'\rangle \langle \beta_2'|.
\end{align}
The joint probability distribution for obtaining quadrature outcomes $x_1$ and $x_2$ is obtained by projecting the detector mode onto the quadrature eigenstates $|x_1\rangle$ and $|x_2\rangle$ and tracing over the field degrees of freedom:
\begin{align}
P(x_1,x_2) &= \mathrm{Tr}_F\left(\langle x_1, x_2 | \rho' | x_1, x_2 \rangle \right)\\
&= \int d^2\alpha \, d^2\beta_1 \, d^2\beta_2 
P(\alpha)\, P(\beta_1)\, P(\beta_2) 
 |\langle x_1 | \beta_1' \rangle|^2 
|\langle x_2 | \beta_2' \rangle|^2.
\end{align}
For a coherent state, the quadrature probability distribution is Gaussian, centered at the real part of the coherent amplitude, and is given by
\begin{align}
|\langle x_i | \beta_i' \rangle|^2
= \frac{1}{\sqrt{\pi}x_{0i}}
\exp\left[
- \frac{\left(x_i - \sqrt{2} x_{0i} \mathrm{Re}(\beta_i') \right)^2}{x_{0i}^2}
\right].
\end{align}
The real part of the transformed modes is directly obtained from Eq.~\eqref{two_det_trans_1} and Eq.~\eqref{two_det_trans_2} and can be written as
\begin{align}
\mathrm{Re}(\beta_1') 
&= \cos^2\left(\frac{\theta_0}{2}\right)\mathrm{Re}(\beta_1)
- \sin^2\left(\frac{\theta_0}{2}\right)\mathrm{Re}(\beta_2)
+ \frac{1}{\sqrt{2}} \sin\theta_0  \mathrm{Im}(\alpha),\label{re_beta1_p}
\\[6pt]
\mathrm{Re}(\beta_2') 
&= \cos^2\left(\frac{\theta_0}{2}\right)\mathrm{Re}(\beta_2)
- \sin^2\left(\frac{\theta_0}{2}\right)\mathrm{Re}(\beta_1)
+ \frac{1}{\sqrt{2}} \sin\theta_0  \mathrm{Im}(\alpha).\label{re_beta2_p}
\end{align}
The resulting joint probability distribution is explicitly written as, 
 \begin{align}
P(x_1,x_2) 
&= \frac{1}{\pi x_{01} x_{02}} 
\int d^2\alpha \, d^2\beta_1 \, d^2\beta_2 
P(\alpha)\, P(\beta_1)P(\beta_2) \\
&\quad \times \exp\left[
- \frac{\left(x_1 - \sqrt{2} x_{01} \mathrm{Re}(\beta_1')\right)^2}{x_{01}^2}
\right]  \exp\left[
- \frac{\left(x_2 - \sqrt{2} x_{02} \mathrm{Re}(\beta_2')\right)^2}{x_{02}^2}
\right].
\end{align}
Once the joint probability distribution is known, the detector correlation can be calculated by evaluating its first and second moments:
\begin{align}
\langle x_1 \rangle &= \sqrt{2} x_{01} 
\langle \mathrm{Re}(\beta_1') \rangle, \quad
\langle x_2 \rangle = \sqrt{2} x_{02} \langle \mathrm{Re}(\beta_2') \rangle,
\label{homo_x1}
\end{align}
and, 
\begin{align}
\langle x_1 x_2 \rangle 
= 2 x_{01} x_{02}  
\langle \mathrm{Re}(\beta_1')\mathrm{Re}(\beta_2') \rangle.\label{homo_x1x2}
\end{align}
Substituting the Eq.~\eqref{re_beta1_p} and Eq.~\eqref{re_beta2_p} into Eq.~\eqref{homo_x1} and Eq.~\eqref{homo_x1x2}, we obtain,
\begin{align}
\langle x_1 x_2 \rangle - \langle x_1 \rangle \langle x_2 \rangle=& 2 x_{01} x_{02}
\Bigg[
\cos^2\frac{\theta_0}{2}
\sin^2\frac{\theta_0}{2}
\left(
\left\langle \mathrm{Re}(\beta_1) \right\rangle^2
-
\left\langle \mathrm{Re}(\beta_1)^2 \right\rangle
\right)
\nonumber\\
&\qquad +
\cos^2\frac{\theta_0}{2}
\sin^2\frac{\theta_0}{2}
\left(
\left\langle \mathrm{Re}(\beta_2) \right\rangle^2
-
\left\langle \mathrm{Re}(\beta_2)^2 \right\rangle
\right)
\nonumber\\
&\qquad +
\frac{1}{2}\sin^2\theta_0
\left(
\left\langle \mathrm{Im}(\alpha)^2 \right\rangle
-
\left\langle \mathrm{Im}(\alpha) \right\rangle^2
\right)
\Bigg].
\end{align}
Here also, we assumed that detectors are initially prepared in an identical, zero-mean state in order to simplify the calculation. Then the above expression can be rewritten in terms of the quadrature variances of the detector and field modes:
\begin{align}
\langle x_1 x_2 \rangle - \langle x_1 \rangle \langle x_2 \rangle=&  2 x_{01} x_{02}
\Bigg[-\frac{1}{2}
\cos^2\frac{\theta_0}{2}
\sin^2\frac{\theta_0}{2} \Big[\langle (\Delta X_1)^2\rangle -\frac{1}{2}\Big]
\nonumber\\
&\qquad -\frac{1}{2}
\cos^2\frac{\theta_0}{2}
\sin^2\frac{\theta_0}{2}
\Big[\langle (\Delta X_2)^2\rangle -\frac{1}{2}\Big]
\nonumber\\
&\qquad +
\frac{1}{4}\sin^2\theta_0 \Big[\langle (\Delta P)^2\rangle -\frac{1}{2}\Big]
\Bigg]. 
\end{align}
Here, $X_1$ and $X_2$ denote the position quadratures of detector 1 and detector 2, respectively, while P corresponds to the momentum quadrature of the field. In the weak coupling limit, the above expression can be further simplified to,
 \begin{align}
\langle x_1 x_2 \rangle
-
\langle x_1\rangle\langle x_2\rangle
\approx
x_{01}x_{02}\theta_0^2
\left[
-\frac{1}{4}\Big[\langle (\Delta X_1)^2\rangle -\frac{1}{2}\Big]
-\frac{1}{4}\Big[\langle (\Delta X_2)^2\rangle -\frac{1}{2}\Big]
+\frac{1}{2}\Big[\langle (\Delta P)^2\rangle -\frac{1}{2}\Big]
\right].
\end{align}
For the two detector model, using $\theta_0=\sqrt{2\gamma_0\tau}$, we obtain, 
\begin{align}
\langle x_1 x_2 \rangle
-
\langle x_1\rangle\langle x_2\rangle
\approx
\frac{\gamma_0\tau}{2}\,
x_{01}x_{02}
\left[
2\Big[\langle (\Delta P)^2\rangle -\frac{1}{2}\Big]
-\Big[\langle (\Delta X_1)^2\rangle -\frac{1}{2}\Big]
-\Big[\langle (\Delta X_2)^2\rangle -\frac{1}{2}\Big]
\right].
\end{align}
This result shows that the correlation in homodyne measurement is proportional to the interaction strength $\gamma_0\tau$ and directly probes the momentum quadrature fluctuations of the field, while subtracting the contribution due to the position quadrature detector noise. Importantly, the quantum mechanical states that show substantial noise are sub-Poissonian quantum states of the radiation field, such as a Fock state, which shows very little noise in the direct counting strategy. Hence Homodyne cross-correlation can provide quantum information that can be complementary to counting detection. This aspect is also highlighted in Refs.~\cite{Manikandan:2025hlz,Manikandan:2025qgv,Manikandan:2025lfx}. In comparison, the additional insight our analysis offers is that a barely functional detector regime offers additional statistical advantages even for thermal preparations of the detector, such that quantum ground state cooling of detectors may not be necessary to probe weak-field quantum effects through homodyne cross-correlation measurements, unlike previously thought.

\section{Correlation in Heterodyne Measurement\label{App_Heterodyne}}
We now consider heterodyne detection of two identical detectors, in which each detector mode is projected onto the coherent state $|\gamma_1\rangle$ and $|\gamma_2\rangle$. The joint probability distribution for obtaining outcomes  $\gamma_1$ and $\gamma_2$ is given by, 
\begin{align}
P(\gamma_1,\gamma_2) &= \frac{1}{\pi^2}\mathrm{Tr}_F\left( \langle \gamma_1, \gamma_2 | \rho' | \gamma_1, \gamma_2 \rangle \right) \nonumber\\
&= \int d^2\alpha \, d^2\beta_1 \, d^2\beta_2 
P(\alpha)\, P(\beta_1)\, P(\beta_2) 
  |\langle \gamma_1 | \beta_1' \rangle|^2 
|\langle \gamma_2 | \beta_2' \rangle|^2.
\end{align}
Using the overlap between coherent states $|\langle \gamma|\beta\rangle|^2
=
e^{-|\gamma-\beta|^2}$, the probability distribution can be written as,
\begin{align}
P(\gamma_1,\gamma_2)
=
\frac{1}{\pi^2}
\int d^2\alpha d^2\beta_1 d^2\beta_2
P_F(\alpha)
P_1(\beta_1)
P_2(\beta_2)
e^{-|\gamma_1-\beta_1'|^2}
e^{-|\gamma_2-\beta_2'|^2}.
\end{align}
The moment of the heterodyne detection is obtained using, 
\begin{align}
\langle \mathrm{Re}(\gamma_i)\rangle
=
\frac{1}{\pi^2}
\int d^2\gamma_1\, d^2\gamma_2
\mathrm{Re}(\gamma_i)
\int d^2\alpha d^2\beta_1 d^2\beta_2
P_F(\alpha)
P_1(\beta_1)
P_2(\beta_2)
e^{-|\gamma_1-\beta_1'|^2}
e^{-|\gamma_2-\beta_2'|^2}.
\end{align}
Here, using, 
\begin{equation}
\frac{1}{\pi}
\int d^2\gamma\;
\mathrm{Re}(\gamma)
e^{-|\gamma-\mu|^2}
=
\mathrm{Re}(\mu),    
\end{equation}
we obtain $\langle \mathrm{Re}(\gamma_1)\rangle
=
\langle \mathrm{Re}(\beta_1')\rangle$, $\langle \mathrm{Re}(\gamma_2)\rangle
=
\langle \mathrm{Re}(\beta_2')\rangle$, and $\langle \mathrm{Re}(\gamma_1) \mathrm{Re}(\gamma_2) \rangle
=\langle \mathrm{Re}(\beta_1') \mathrm{Re}(\beta_2') \rangle$.
Therefore, the correlation in the measured real part is identical to the correlation in transformed detector amplitudes:
\begin{align}
\langle
\mathrm{Re}(\gamma_1)
\mathrm{Re}(\gamma_2)
\rangle
-
\langle \mathrm{Re}(\gamma_1)\rangle
\langle \mathrm{Re}(\gamma_2)\rangle
\nonumber
=&
\langle
\mathrm{Re}(\beta_1')
\mathrm{Re}(\beta_2')
\rangle
-
\langle \mathrm{Re}(\beta_1')\rangle
\langle \mathrm{Re}(\beta_2')\rangle\\
=&
\cos^2\frac{\theta_0}{2}
\sin^2\frac{\theta_0}{2}
\left(
\langle \mathrm{Re}(\beta_1)\rangle^2
-
\langle \mathrm{Re}(\beta_1)^2\rangle
\right)
\nonumber\\
&+
\cos^2\frac{\theta_0}{2}
\sin^2\frac{\theta_0}{2}
\left(
\langle \mathrm{Re}(\beta_2)\rangle^2
-
\langle \mathrm{Re}(\beta_2)^2\rangle
\right)
\nonumber\\
&+
\frac{1}{2}\sin^2\theta_0
\left(
\langle \mathrm{Im}(\alpha)^2\rangle
-
\langle \mathrm{Im}(\alpha)\rangle^2
\right).
\end{align}
This expression can be further simplified by writing in terms of the quadrature variances of the detector and field modes, as in the Homodyne case:
\begin{align}
\langle
\mathrm{Re}(\beta_1')
\mathrm{Re}(\beta_2')
\rangle
-
\langle \mathrm{Re}(\beta_1')\rangle
\langle \mathrm{Re}(\beta_2')\rangle=&  
\Bigg[-\frac{1}{2}
\cos^2\frac{\theta_0}{2}
\sin^2\frac{\theta_0}{2} \Big[\langle (\Delta X_1)^2\rangle -\frac{1}{2}\Big]
\nonumber\\
&\qquad -\frac{1}{2}
\cos^2\frac{\theta_0}{2}
\sin^2\frac{\theta_0}{2}
\Big[\langle (\Delta X_2)^2\rangle -\frac{1}{2}\Big]
\nonumber\\
&\qquad +
\frac{1}{4}\sin^2\theta_0 \Big[\langle (\Delta P)^2\rangle -\frac{1}{2}\Big]
\Bigg].
\end{align}
In the weak coupling limit, this expression reduces to
\begin{align}
\mathrm{Re}(\beta_1')
\mathrm{Re}(\beta_2')
\rangle
-
\langle \mathrm{Re}(\beta_1')\rangle
\langle \mathrm{Re}(\beta_2')\rangle
\approx
\frac{\gamma_0\tau}{2}
\left[
\Big[\langle (\Delta P)^2\rangle -\frac{1}{2}\Big]
-\frac{1}{2}\Big[\langle (\Delta X_1)^2\rangle -\frac{1}{2}\Big]
-\frac{1}{2}\Big[\langle (\Delta X_2)^2\rangle -\frac{1}{2}\Big]
\right].
\end{align}
Here, compared with homodyne detection, heterodyne detection gives half of the leading-order contribution from both the field quadrature fluctuations and the detector fluctuations, but otherwise the conclusions drawn are in agreement with the previous sessions. 
\section{Connection to Bolometric Detection Strategies\label{App_bolo}}
Here, we briefly summarize the links to bolometric detection strategies in quantum optics, which we take inspiration from. We explain the idea using the simplest, single-detector model explained in the main text and in Appendix.~\ref{Appx-A}. This can be extended to more elaborate models of a solid using the normal mode analysis presented in the manuscript. We also consider an incoming thermal radiation field interacting with a detector initialized in its thermal state. After the interaction, the states of the detector become, 
\begin{equation}
\rho_D' 
= \int d^2\alpha \, d^2\beta \;
P_{F}(\alpha)\, P_{D}(\beta) \ket{\beta'}\bra{\beta'},
\end{equation}
where following from Appendix.~\ref{Appx-A},
\begin{align}
\alpha' &= \alpha\cos(\sqrt{\gamma_0 \tau}) - i\beta\sin(\sqrt{\gamma_0 \tau}),
\label{1d_trans_1}\\
\beta'  &= \beta\cos(\sqrt{\gamma_0 \tau}) - i\alpha\sin(\sqrt{\gamma_0 \tau}).
\label{1d_trans_2}
\end{align}
For both the thermal radiation field and the detector, the Glauber-Sudarshan $P$ function is given by,
\begin{align}
    P(\alpha) = \frac{1}{\pi\, \bar{n}_{F/D}} \exp\left[-\frac{|\alpha|^2}{\bar{n}_{F/D}}\right],
\end{align}
where $\bar{n}_{F/D}$ denotes the mean number of thermal quanta in the field/detector, respectively. Using this, the state of the detector after the interaction, and tracing over the field, becomes,
\begin{equation}
\rho_D' 
= \frac{1}{\pi^2\,\bar{n}_F\,\bar{n}_D}\int d^2\alpha' \, d^2\beta' \;
\exp\left[-\frac{|\alpha|^2}{\bar{n}_F}\right]\, \exp\left[-\frac{|\beta|^2}{\bar{n}_D}\right] \ket{\beta'}\bra{\beta'},
\label{1d_detector_den}
\end{equation}
where we have transformed the integration variable form $\alpha,\beta$ to $\alpha', \beta'$. This change of variables is allowed because the transformations given in Eq.~\eqref{1d_trans_1} and Eq.~\eqref{1d_trans_2} are unitary. Hence, the phase space volume does not change under this transformation, and the Jacobian determinant is unity. Now, by inverting Eq.~\eqref{1d_trans_1} and Eq.~\eqref{1d_trans_2} and substituting the resulting expressions into Eq.~\eqref{1d_detector_den}, the resulting integral can be written entirely in terms of $\alpha'$ and $\beta'$. Here, the integration over $\alpha'$ can be done in a straightforward way, as it is a general Gaussian integral, which would give the result

\begin{equation}
\begin{aligned}
\rho_D' 
&= \frac{1}{\pi^2 \bar{n}_F \bar{n}_D}
\int d^2\beta' \;
\frac{
2\pi \bar{n}_D \bar{n}_F
\exp\!\left[
-\dfrac{|\beta'|^2}
{\bar{n}_D \cos^2\theta + \bar{n}_F \sin^2\theta}
\right]
}
{
\cos(2\theta)(\bar{n}_D-\bar{n}_F)
+\bar{n}_D+\bar{n}_F
}
\;
\ket{\beta'}\bra{\beta'} , \quad  \text{assuming} \quad \theta = \sqrt{\gamma_0 \tau}.
\end{aligned}
\end{equation}
After simplification, we get,
\begin{equation}
\begin{aligned}
\rho_D'
=
\frac{1}{\pi\left(
\bar{n}_D \cos^2\theta+
\bar{n}_F \sin^2\theta
\right)}
\int d^2\beta'\,
\exp\!\left[
-\frac{|\beta'|^2}
{
\bar{n}_D \cos^2\theta
+
\bar{n}_F \sin^2\theta
}
\right]
\ket{\beta'}\bra{\beta'}.
\end{aligned}
\end{equation}
We note that, statistically, the new detector state is again a thermal state with an effective average occupation number $\bar{n}_{\mathrm{eff}} = \bar{n}_D \cos^2\theta + \bar{n}_F \sin^2\theta $. We can also associate an effective temperature through the Bose-Einstein statistics, 
\begin{equation}
    \bar{n}_{\mathrm{eff}} = \frac{1}{\exp\left[\frac{\hbar\,\omega}{k_B\,T_{\mathrm{eff}}}\right]-1}.
\end{equation}
Solving for the new effective temperature of the detector $T_{\mathrm{eff}}$ gives,
\begin{equation}
 T_{\mathrm{eff}} = \frac{\hbar\omega}
{k_B \ln\left[1+\frac{1}{\bar{n}_{\mathrm{eff}}}\right]} =\frac{\hbar\omega}
{k_B \ln\left[1+\frac{1}{\bar{n}_D \cos^2\theta + \bar{n}_F \sin^2\theta}\right]}  .   
\end{equation}
Bolometry ideas indeed use measurement of temperature, and temperature fluctuations as probes for the radiation field, and our analysis here suggests that the results presented in the main text can also be cast in the bolometric framework, systematically. We defer a detailed analysis of these to future work.

Similarly, if we calculate the state of the radiation field after interaction, we can show that it is also a thermal state with an effective occupation number $\bar{n}_{\mathrm{F_{eff}}} = \bar{n}_F \cos^2\theta +\bar{n}_D \sin^2\theta $. The tracing over the other degree of freedom serves as a refresh protocol, and if one continues the process in sequence, iteratively, the population dynamics of both the detector and field would correspond to thermal relaxation of a system of two coupled quantum harmonic oscillators, discussed in Ref.~\cite{Chimonidou:2007nc}.

\end{widetext}
\bibliography{citation}
\end{document}